# Identification of interictal epileptic networks from dense-EEG

Mahmoud Hassan[1,2], Isabelle Merlet[1,2], Ahmad Mheich[1,2,4], Aya Kabbara[1,2,4], Arnaud Biraben[1,2,3], Anca Nica[3] and Fabrice Wendling[1,2]

[1] INSERM, U1099, Rennes, F-35000, France

[2] Université de Rennes 1, LTSI, F-35000, France

[3] Neurology department, CHU, Rennes, France

[4] AZM center-EDST, Lebanese University, Tripoli, Lebanon

Corresponding author:

Mahmoud Hassan

mahmoud.hassan@univ-rennes1.fr

Tel: +33 2 23 23 56 05 / 62 20

Fax: +33 2 23 23 69 17

# Abstract

Epilepsy is a network disease. The epileptic network usually involves spatially distributed brain regions. In this context, noninvasive M/EEG source connectivity is an emerging technique to identify functional brain networks at cortical level from noninvasive recordings. In this paper, we analyze the effect of the two key factors involved in EEG source connectivity processing: i) the algorithm used in the solution of the EEG inverse problem and ii) the method used in the estimation of the functional connectivity. We evaluate four inverse solutions algorithms (dSPM, wMNE, sLORETA and cMEM) and four connectivity measures ($r^2$, $h^2$, PLV, and MI) on data simulated from a combined biophysical/physiological model to generate realistic interictal epileptic spikes reflected in scalp EEG. We use a new network-based similarity index (SI) to compare between the network identified by each of the inverse/connectivity combination and the original network generated in the model. The method will be also applied on real data recorded from one epileptic patient who underwent a full presurgical evaluation for drug-resistant focal epilepsy.

In simulated data, results revealed that the selection of the inverse/connectivity combination has a significant impact on the identified networks. Results suggested that nonlinear methods (nonlinear correlation coefficient, phase synchronization and mutual information) for measuring the connectivity are more efficient than the linear one (the cross correlation coefficient). The wMNE inverse solution showed higher performance than dSPM, cMEM and sLORETA. In real data, the combination (wMNE/PLV) led to a very good matching between the interictal epileptic network identified from noninvasive EEG recordings and the network obtained from connectivity analysis of intracerebral EEG recordings. These results suggest that source connectivity method, when appropriately configured, is able to extract highly relevant diagnostic information about networks involved in interictal epileptic spikes from non-invasive dense-EEG data.

# Introduction

Epilepsy is a network disease (Engel Jr et al. 2013). Over the two past decades, the concept of "epileptic focus" has progressively evolved toward that of "epileptic network" (Kramer and Cash 2012; Laufs 2012). Indeed, with the progress of functional neuroimaging techniques, many studies confirmed that the epileptic zone (EZ) can rarely be reduced to a circumscribed brain area (Bartolomei et al. 2001) as it very often involves distinct brain regions generating both interictal (Bourien et al. 2005) and ictal activity (Bourien et al. 2004). Among the investigation techniques classically used in the diagnostic of epilepsy, electrophysiological recordings (typically, magneto- and electro-encephalography including depth-EEG, referred to as M/EEG) are still extensively used to localize and delineate the EZ in a patient-specific context. Regarding the numerous methods proposed to process the recorded data; those aimed at characterizing brain connectivity are particularly suitable to identify networks involved during epileptiform activity (both interictal and ictal).

In the context of invasive EEG signals (intracranial EEG, stereo-EEG and electrocorticoGraphy –EcoG-) recorded in patients candidate to surgery, these "connectivity" methods have been a topic of extensive research (see (van Mierlo et al. 2014) for recent review). For instance, the coherence function was shown to localize the seizure onset (Gotman 1987), similarity indexes were used to distinguish a preictal state from the ongoing interictal activity (Le Van Quyen et al. 2005; Mormann et al. 2000). Nonlinear regression analysis was applied to intracerebral signals to characterize connectivity patterns at the seizure onset (Bourien et al. 2004). Readers may refer to previous reviews for more detailed information about brain connectivity methods applied to non-invasive (van Mierlo et al. 2014) and invasive EEG signals (Wendling et al. 2010) in drug-resistant focal epilepsies.

In the context of scalp M/EEG recording, connectivity methods have received less attention as compared with invasive EEG. A number of studies performed at the level of electrodes and focused on ictal periods have been reported aiming at analyzing seizure propagation (Gotman 1983) or to determine the seizure onset side (Caparos et al. 2006), for instance. For interictal periods, few connectivity studies made use of dense EEG and phase synchronization (Ramon and Holmes 2013) to identify epileptic sites. One reason for this paucity of studies may lie in the intricate interpretation of connectivity measures obtained from scalp recordings. Indeed, this interpretation is not straightforward as signals are severely corrupted by the effects of volume conduction (Schoffelen and Gross 2009).

Interestingly, some recent studies showed how to overcome this limitation. In line with previous cognitive studies (Astolfi et al. 2007; Babiloni et al. 2005; Betti et al. 2013; Bola and Sabel 2015; David et al. 2003; David et al. 2002; de Pasquale et al. 2010; Hassan et al. 2015a; Hassan et al. 2014; Hassan and Wendling 2015; Hipp et al. 2011; Hoechstetter et al. 2004; Liljeström et al. 2015; Schoffelen and Gross 2009), the basic principle is to estimate functional connectivity at the level of brain sources reconstructed from scalp signals. These methods, referred to as "source connectivity" were applied to both interictal EEG (Coito et al. 2015; Song et al. 2013; Vecchio et al. 2014) and MEG signals (Dai et al. 2012; Malinowska et al. 2014) as well as to EEG signals recorded during seizures (Ding et al. 2007; Jiruska et al. 2013; Lu et al. 2012) or resting states (Adebimpe et al. 2016; Coito et al. 2016).

Although these approaches all include two steps (M/EEG inverse problem followed by source connectivity estimation), they strongly differ from a methodological viewpoint. Indeed, various algorithms were used to reconstruct brain sources and both functional and effective connectivity measures were utilized to assess statistical couplings among time series associated with reconstructed sources. Therefore, a central issue is the impact of selected methods (EEG inverse solution and connectivity measure) on the topological/statistical properties of identified epileptic networks activated during paroxysmal activity.

In this paper, we report a quantitative comparison of methods aimed at identifying cortical epileptic networks from scalp EEG data. The novelty of this work is twofold. First, our comparative study includes simulated dense EEGs generated from physiologically- and biophysically-plausible models of distributed and coupled epileptic sources. To our knowledge, no previous study has reported results on the performance of source connectivity methods based on a "ground truth" provided by realistic computational models of interictal EEG signals (recorded later in time than the dense EEG recordings). Second, in line with a recent analysis performed on MEG data (Malinowska et al. 2014), networks estimated from real scalp dense EEG are compared with those obtained from depth-EEG recordings (SEEG).

# Materials and Methods

*A. Inverse solution algorithms*

Given the equivalent current dipole model, EEG signals X(t) recorded from M channels can be considered as linear combinations of P time-varying current dipole sources S(t):

$$X(t) = GS(t) + N(t)$$

where G[M, P] is the lead field matrix and N(t) is the noise. As G is known, the EEG inverse problem consists of estimating the unknown sources $\hat{S}(t)$ from X(t). Several algorithms have been proposed to solve this problem based on different assumptions about spatial and temporal properties of sources and regularization constraints. Here, we chose to evaluate the four algorithms implemented in Brainstorm (Tadel et al. 2011).

*1) Weighted Minimum Norm Estimate (wMNE)*

Minimum norm estimates (MNE) originally proposed by (Hämäläinen and Ilmoniemi 1994) are based on a search for the solution with minimum power using the L2 norm to regularize the problem. A common expression for MNE resolution

matrix is

$$\hat{S}_{MNE} = G^T(GG^T + \lambda C)^{-1}G$$

where $\lambda$ is the regularization parameter and C represents the noise covariance matrix. The weighted MNE (wMNE) algorithm compensates for the tendency of MNE to favor weak and surface sources (Hämäläinen 2005). This is achieved by introducing a weighting matrix $W_X$ in

$$\hat{S}_{wMNE} = (G^T W_X G + \lambda C)^{-1} G^T W_X X$$

that adjusts the properties of the solution by reducing the bias inherent to the standard MNE solution. Classically, $W_X$ is a diagonal matrix built from matrix G with non-zero terms inversely proportional to the norm of the lead field vectors.

*2) dynamical Statistical Parametric Mapping (dSPM)*

The dSPM is based on the MNE solution (Dale et al. 2000). For dSPM, the normalization matrix contains the minimum norm estimates of the noise at each source (Caparos et al. 2006), derived from the noise covariance matrix, defined as:

$$\hat{S}_{dSPM} = W_{dSPM} \hat{S}_{MNE}$$

Where $W_{dSPM}^2 = \text{diag}(\hat{S}_{MNE} C \hat{S}_{MNE}^T)$.

*3) Standardized low resolution brain electromagnetic tomography (sLORETA)*

sLORETA uses the source distribution estimated from MNE and standardizes it a posteriori by the variance of each estimated dipole source:

$$\hat{S}_{sLORETA} = W_{sLORETA} \hat{S}_{MNE}$$

where $W_{sLORETA}^2 = \text{diag}(\hat{S}_{MNE} G) = \text{diag}(\hat{S}_{MNE}(GG^T + C)\hat{S}_{MNE}^T)$. The sLORETA inverse method has been originally described using the whole brain volume as source space (Pascual-Marqui 2002). For the present study, in order to ease the comparison with other methods, we restricted the source space to the neocortical surface.

*4) Standard Maximum Entropy on the Mean (cMEM)*

The Maximum Entropy on the Mean (MEM) solver is based on a probabilistic method where inference on the current source intensities is estimated from the data, which is the basic idea of the maximum of entropy. The first application of MEM to electromagnetic source localization was reported in (Clarke and Janday 1989). The main feature of this method is its ability to recover the spatial extent of the underlying sources. Its solution is assessed by finding the closest distribution of source intensities to a reference distribution in which source intensities are organized into cortical parcels showing homogeneous activation state (parallel cortical macro-columns with synchronized activity). In addition a constraint of local spatial smoothness in each parcel can be introduced (Chowdhury et al. 2013).

*B. Connectivity measures*

We selected four methods that have been widely used to estimate functional brain connectivity from electrophysiological signals (local field potentials, depth-EEG or EEG/MEG) (see (Wendling et al. 2009) for review). These measures were chosen to cover the main families of connectivity methods (linear and nonlinear regression, phase synchronization and mutual information).

Briefly, concerning the regression approaches, the linear cross-correlation coefficient is only limited to the detection of the linear properties of the relationships between time series. However, mechanisms at the origin of EEG signals were shown to have strong nonlinear behaviors (Pereda et al. 2005). Thus, we have selected three nonlinear connectivity measures. The nonlinear regression where the basic idea is to evaluate the dependency of two signals from signal samples only and independently of the type of relation between the two signals. Concerning the phase synchronization measure, the method estimates the instantaneous phase of each signal and then computes a quantity based on co-variation of extracted phases to determine the degree of relationship. Finally, the mutual information method is based on the probability and information theory to measures mutual dependence between two variables. More technical details about the four methods are presented hereafter:

*1) Cross-correlation coefficient ($r^2$)*

The cross-correlation coefficient measures the linear correlation between two variables *x* and *y* as a function of their time delay ($\tau$). Referred to as the linear correlation coefficient, it is defined as:

$$r_{xy}^2 = \max_{\tau} \frac{\text{cov}^2(x(t), y(t+\tau))}{(\sigma_{x(t)}\sigma_{y(t+\tau)})^2}$$

where $\sigma$ and cov denote the standard deviation and the covariance, respectively.

*2) Nonlinear correlation coefficient ($h^2$)*

The nonlinear correlation coefficient ($h^2$) is a non-parametric measure of the nonlinear relationship between two time series *x* and *y*. In practice, the nonlinear relation between the two time series is approximated by a piecewise linear curve.

$$h_{xy}^2 = \max_{\tau}\left(1 - \frac{\text{var}(y(t+\tau)/x(t))}{\text{var}(y(t+\tau))}\right)$$

where $\text{var}(y(t+\tau)/x(t)) \triangleq \arg\min_{f}\left(E[y(t+\tau) - f(x(t))]^2\right)$ and $f(x)$ is the linear piecewise approximation of the nonlinear regression curve.

*3) Mutual information (MI)*

The mutual information (*MI*) between signal *x* and *y* is defined as:

$$MI_{xy} = \sum p_{ij}^{xy} \log \frac{p_{ij}^{xy}}{p_i^x p_j^y}$$

where $p_{ij}^{xy}$ is the joint probability of $x=x_i$ and $y=y_j$. In the case of no relationship between *x* and *y*, $p_{ij}^{xy} = p_i^x p_j^y$, so that the *MI* is zero for independent processes. Otherwise, $MI_{xy}$ will be positive, attaining its maximal value for identical signals.

*4) Phase Locking Value (PLV)*

For two signals *x* and *y*, the phase locking value is defined as:

$$PLV_{xy} = \left|\langle e^{i|\varphi_x(t) - \varphi_y(t)|}\rangle\right|$$

where $\varphi_x(t)$ and $\varphi_y(t)$ are the unwrapped phases of the signals *x* and *y* at time *t*. The $\langle.\rangle$ denotes the average over time. The Hilbert transform was used to extract the instantaneous phase of each signal.

The $h^2$, PLV and $r^2$ values range from 0 (independent signals) to 1 (fully correlated signals).

*C. Data*

  *Simulations*

In order to quantitatively assess the performance of source connectivity approaches, we generated simulated EEG signals following the procedure described in (Cosandier-Rimélé et al. 2008), see figure 1A. The distributed source space consisted in a mesh of the cortical surface (8000 vertices, ~5 mm inter-vertex spacing) that was obtained by segmenting the grey-white matter interface from a normal subject's structural T1-weighted 3D-MRI with Freesurfer (Fischl 2012). Dipoles were located at each vertex of this mesh and oriented radially to the surface at the midway between the white/grey matter interface and the pial surface. The time-course of each dipole of the source space was generated from a modified version of the physiologically relevant neural mass model reported in (Bourien et al. 2005; Wendling et al. 2002; Wendling et al. 2000).

In brief, this computational model was designed to represent a neuronal population with three subsets of neurons (pyramidal cells P and interneurons I and I') interacting via synaptic transmission (Figure 1-A). Pyramidal cells (P) receive endogenous excitatory drive (AMPAergic collateral excitation) from other local pyramidal cells and exogenous excitatory drive from distant pyramidal cells (via external noise input p(t)). They also receive inhibitory drive (GABAergic feedback inhibition) from both subsets of local interneurons (I and I'). In turn, interneurons receive excitatory input (AMPA) from pyramidal cells.

A Gaussian noise was used as external input to neuronal population. The mean (m=90) and standard deviation (sigma=30) were adjusted to represent randomly varying density of incoming action potentials (Aps). However, for the purpose of this study, a modification was made to this noise model. Indeed, abrupt increase/decrease of the density of Aps can occur in the external input noise at user-defined times to mimic transient AP volleys from other brain regions involved in the generation of interictal events. Thus, in this model, a simulated IES can be viewed as the responses of a nonlinear dynamical system (comprising positive and negative feedback loops) to transient pulses superimposed on a Gaussian noise (classically used in the neural mass modeling approachs).

As in the standard implementation, the shape (spike component followed by a wave component) can still be controlled by adjusting excitation and inhibition parameters of each population (gains in feedback loops). However, the aforementioned modification offers one major advantage: as pulses in the noise input are user-defined, the occurrence times of simulated IESs are controlled, in contrast with the standard implementation where IESs simply result from random fluctuations of the noise. The consequence is that this new model feature allows for simulation multi-focal IESs with well-controlled time shifts. Indeed, as illustrated in Figure 1-A, we could generate delayed epileptiform activity in multiple distant patches just by introducing short delays between the pulses superimposed on the respective input noises of neuronal populations at each patch.

Finally, from appropriate setting of the input noise, as well as excitation and inhibition-related parameters at each neural mass included in simulated epileptic sources, a set of epileptiform temporal dynamics was obtained. These dynamics were assigned to a source made of contiguous vertices (active source) manually outlined with a mesh visualization software (Paraview, Kitware Inc., NY, US). Uncorrelated background activities were attributed to the other vertices. Once the amplitude of each elementary dipole was known, EEG simulations were obtained by solving the forward problem in a 3-layer realistic head model (inner skull, outer skull and the scalp with conductivity values of 0.33, 0.0042, 0.33 S/m respectively) using the Boundary Element Method (BEM) with the OpenMEEG (Gramfort et al. 2010) implemented in Brainstorm software.

We considered two different scenarios. In the first one (single network), EEG simulations were generated from a single source located in the inferior parietal region (~1000 mm²). In the second one (two interconnected networks) an additional source (~1000 mm²) was placed in the middle temporal gyrus. In that case, the temporal dynamics of the second source were highly correlated with those of the first source, but with a minor delay (30ms). This delay of 30 ms was in the range of 10-50 ms delays that are often observed during interictal spikes at different intracranial recording location (Alarcon et al. 1994; Alarcon et al. 1997; Emerson et al. 1995; Merlet and Gotman 1999) or at different surface sensors (Barth et al. 1984; Ebersole 1994) or between the peaks of dipole source activity (Baumgartner et al. 1995; Merlet and Gotman 1999). This delay was usually interpreted as reflecting propagation between distant regions in the brain. For each scenario, 20 epochs of 60s at 512 Hz containing 30 epileptic spikes were simulated. Each epoch was obtained for a new realization of the input random noise leading to a new realization of epileptic spikes occurring in background activity. Simulated data were imported in Brainstorm for further analysis.

*Real data*

Real data were selected from a patient who underwent presurgical evaluation for drug-resistant focal epilepsy. Seizures were stereotyped, with a sudden start, febrile motor automatisms of the upper limb, stretching of the neck and torso and no post-ictal motor deficit. The patient had a comprehensive evaluation including detailed history and neurological examination, neuropsychological testing, structural MRI, standard 32-channels (Micromed, Italy) as well as Dense-EEG 256-channels (EGI, Electrical Geodesic Inc., Eugene, USA) scalp EEG with video recordings and intracerebral EEG recordings (SEEG). MRI showed a focal cortical dysplasia in the mesial aspect of the orbito-frontal region. Dense-EEG was recorded for one hour, at 1000 Hz following the procedure approved by the National Ethics Committee for the Protection of Persons (CPP, agreement number 2012-A01227-36). The patient gave his written informed consent to participate in this study. This recording revealed sub-continuous spike activity at the most left frontopolar basal electrodes. From this interictal epileptic activity, 85 spikes were visually selected away from the occurrence of any artefacts (muscle activity, blood pulsation, eye blinks). Each spike was centered in a 2s window and all 85 windows were concatenated for further analysis.

As part of his presurgical evaluation, the patient also underwent intracerebral SEEG recordings with 9 implanted electrodes (10±18 contacts; length: 2 mm, diameter: 0.8 mm; 1.5 mm apart) placed intracranially according to Talairach's stereotactic method in the left frontal and temporal region, see Figure 1C. The positioning of the electrodes was determined from available non-invasive information and hypotheses about the localization of his epileptic zone. From these data, subsets of 25 out of the 118 original leads were selected. This selection was made according to the following criteria: i) only contacts showing grey matter activity were retained and ii) among them, only the contact showing the maximal activity was kept when similar intracerebral activity was observed on several contacts.

*D. Data analysis*

*Scalp-EEG based interictal epileptic networks.* As illustrated in Figure 1B, source activity was estimated using four inverse algorithms (dSPM, wMNE, sLORETA and cMEM, see *materials and methods* section –A). A baseline of 1s length was used to estimate the noise covariance matrix both on simulated and real scalp EEG data. For real data, source localization was applied on averaged spikes, taking as time reference the maximum of the negative peak, while for simulated data the source localization was applied to non-averaged spikes. The cortical surface was anatomically parcellated into 148 regions of interest (ROI) (Destrieux et al. 2010) and then re-subdivided into ~1500 sub-ROIs using Brainstorm. The 148 ROIs provided initially by the Destrieux Atlas (using Freesurfer) were *quasi* equally subdivided to obtain the 1500 sub-ROIs with 1cm$^2$ average sizes. Time series of the reconstructed source activities were averaged over each of the 1500 ROIs.

After the reconstruction of sources (source localization and estimation of temporal dynamics), functional connectivity was estimated using four methods ($r^2$, $h^2$, PLV, and MI, see *materials and methods* section -A-). Each quantity was computed on the set of 2 s single spikes. All connectivity matrices (1500 x 1500) were thresholded as follows. We computed the strength of each node of the weighted undirected graph and we kept nodes with the highest 1% strength values. The same threshold was applied on the adjacency matrices for all combinations (inverse/connectivity). The strength was defined as the sum of all edge weights for each node; all weights were positive and normalized between 0 and 1.

In order to define the reference networks, all the dipoles were supposed synchronized and the reference network reflected a fully connected undirected graph. In the case of double network scenario, a number of 37 sub-regions (nodes) were considered. The dynamics of the dipoles associated to these nodes were similar and resulting a 37x37 fully connected network where connections (local and remote) between the 37 nodes have the same weight value.

*Quantification of network similarity.* In order to compare the reference brain network simulated in the model with the network identified from simulated scalp EEG by each of the inverse/connectivity combination (Figure 1B), we used a network similarity algorithm recently developed in our team (Mheich et al. 2015a), see *supplementary materials* for more details about the algorithm. The main advantage of this algorithm is that it takes into account the spatial location (3D coordinates) of the nodes when comparing two networks, in contrast with other methods based on the sole statistical properties of compared graphs. The algorithm provides a normalized Similarity Index (SI): 0 for no similarity and 1 for two identical networks (same properties and topology). The connectivity analysis, the network measures and network visualization were performed using EEGNET (Hassan et al. 2015a; Hassan et al. 2015b).

*Depth-EEG based interictal epileptic networks.* Functional connectivity using $h^2$ were directly computed from SEEG signals at the 25 selected intracerebral electrode contacts. Adjacency matrices (25 x 25) were obtained and thresholded using the same procedure than that applied to the graphs obtained for scalp dense EEG (both simulated and real).

*Scalp-EEG-based vs. depth-EEG-based epileptic network matching.* In order to compare the graphs in the three-dimensional coordinates system of the cortex mesh, the 3D coordinates of the SEEG were first estimated by the co-registering the patient' CT scan and MRI. These points were then projected on the surface mesh. The transformation from MRIs (voxels) coordinates to surface (SCS/MNI) coordinates was realized in brainstorm. The Scalp-EEG-based and depth-EEG-based epileptic networks were visually compared by matching the identified regions (nodes) in both networks.

*Statistical analysis:* On the simulated data, a Wilcoxon rank-sum test was used to compare between the SIs obtained for each combination at each trial, corrected for multiple comparison using Bonferroni approach.

# Results

*A. Simulated Data: Influence of the source reconstruction/functional connectivity combination*

The results obtained in the case of the single network scenario are illustrated in Figure 2, for the 16 different combinations of the source reconstruction and functional connectivity methods. The visual investigation of these results revealed that networks identified using the different combinations of methods were concordant with the reference network (figure 2B). Indeed none of the identified networks had nodes in a remote region (Figure 2A). The qualitative analysis also showed that the number of nodes and the connections between them varied according to the combination of methods used. For a given connectivity approach, changing the localization method did not dramatically modify the network aspect, except for cMEM. Conversely, for a given source localization approach, changing the functional connectivity measure changed, qualitatively, the network. Although this was difficult to assess visually, $h^2$ combined with MNE or sLORETA was giving the network that best matched the reference network while cMEM/MI provided a result that was different from the reference network. Quantification of these differences is provided in figure 2C. Overall, values of network similarity were relatively high and ranged from 70 to 82%. For a given connectivity approach, changing only the localization algorithm slightly modified SI values by 3% ($h^2$) to 8% (MI). For a given source localization approach, the SIs varied within 9% (wMNE) to 12% (dSPM). Results obtained using MI were on average better than PLV, $r^2$ and $h^2$. The combination providing the highest similarity values between the estimated and the actual network was dSPM/MI (82.2%) followed by wMNE/MI (82%) and wMNE-PLV (82%). The lowest similarity value was obtained with the dSPM/$h^2$ combination. The Wilcoxon rank-sum test did not reveal any significant difference between the similarity values obtained in this first study.

Results obtained in the case of two interconnected networks for the 16 combinations of the inverse/connectivity methods are reported in Figure 3. Results indicate that the networks identified by all the combinations are relatively close to the model network (figure 3B) since, similarly to the previously scenario, there was no node in other distant regions or in the right hemisphere. The networks did not qualitatively change much for a given connectivity measure except for cMEM. Rather, as observed in the first scenario, the variability between the different combinations was more related to the choice of the connectivity measure, given a source localization approach. The results of PLV (whatever the inverse solution algorithm) provide the closest result to the reference network. cMEM/MI shows also a relatively close network to the reference network while cMEM/$h^2$ indicated, visually, the farthest result from the reference network.

Values of network similarity are reported in figure 3C. These values were lower than those in the single network scenario, ranging from 57 to 73%. For a particular connectivity measure, changing the inverse algorithm modified the SIs by 1% ($r^2$) to 8% ($h^2$) while for a given source reconstruction algorithm, the SIs varied between 6% (dSPM) to 13% (wMNE). The combination providing the result closest to the reference network was wMNE/PLV (73%). High values were also obtained with sLORETA/PLV (68%) and cMEM/PLV (66%). The cMEM/ $h^2$ combination shows the lowest SI value (57%).

Interestingly, for scenario 2 results obtained with wMNE/PLV were significantly closest to the actual network than the other ones (Wilcoxon rank-sum test, $p<0.01$, corrected using Bonferroni).

### B. EEG source localization vs. functional connectivity

An essential issue that is addressed in this paper relates to the difference between the proposed "network-based" approach and the classical approach using source localization only. In figure 4, we show two typical examples of the difference between the proposed network-based analysis and the classical localization approach. The first example is for cMEM combined with MI vs. cMEM only. This Figure shows that the information extracted in both cases was noticeably different. The source connectivity approach identified a network close to the reference one (figure 4A), with nodes both in the parietal and in the temporal region (figure 4B). There were no spurious nodes in remote regions. In contrast, with the sole source localization, after averaging the results over a 50 ms interval around each of the epileptic peaks, the parietal source was well retrieved while the temporal source remained almost unobserved. The second example was wMNE/PLV vs. wMNE, the figure shows that the network-based approach was able to identify a network close to the reference with no spurious connections in distant regions. The source localization approach identified the two regions different energies at. Moreover, many spurious sources were observed in remote regions. Similar results were observed for single network configuration.

### C. Real data: Scalp-EEG-based vs. depth-EEG-based epileptic network

The results obtained from real data recorded in a patient are described on Figure 5. In this patient, the sources of scalp EEG interictal spikes were widespread over the left frontal and temporal regions. Sources with maximum activity were found in the left frontal pole and orbitofrontal regions but a substantial activation was also detected in the left temporal as well as right frontal poles (figure 5A, left). When combining wMNE and PLV on the same scalp EEG data, the source connectivity approach retrieved a 5-nodes network in the left frontal lobe, involving the mesial (rectus gyrus) and lateral orbitofrontal region as well as the anterior cingulate gyrus (figure 5A, right). This result was concordant with that the network identified directly from intracerebral recordings by computing the functional connectivity among SEEG signals (Figure 5B right). Indeed, the depth-EEG based network involved six nodes in the left mesial orbito-frontal (rectus gyrus), and anterior cingulate region. All these nodes were also identified by the visual analysis (Figure 5B, left) as regions involved in the main interictal activity (rectus gyrus) as well as in the propagated interictal activity (cingulate gyrus).

The similarity indices between networks identified by each of the combination with the depth-EEG-based network are presented in figure 5C. Results showed that the wMNE/PLV provides the highest SI value (70%) followed by wMNE/$h^2$ (47%) and sLORETA/PLV (47%). The cMEM method showed the lowest SI values whatever the connectivity measure (6%, 1%, 1% and 1% for cMEM/MI, cMEM/PLV, cMEM/$h^2$ and cMEM/$r^2$ respectively).

# Discussion

Identifying epileptic brain networks from noninvasive M/EEG data is a challenging issue. Recently, source localization combined with functional connectivity analysis led to encouraging findings in the estimation of functional cortical brain networks from scalp M/EEG recordings (Coito et al. 2015; Jiruska et al. 2013; Malinowska et al. 2014). Nevertheless, the joint use of these two approaches is still novel and raises a number of methodological issues that should be controlled in order to get appropriate and interpretable results. In this paper, we reported a comparative study -in the context of epilepsy- of the networks obtained from all possible combinations between four algorithms to solve the EEG inverse problem and four methods to estimate the functional connectivity. An originality of this study is related to the use of dense EEG signals simulated data from a realistic model of multi-focal epileptic zone as a ground truth for comparing the performance of considered methods. To our knowledge, a model-based evaluation of source connectivity methods has not been performed yet. A second – and still novel - aspect is the use of depth-EEG signals (intracerebral recordings performed during presurgical evaluation of drug-resistant epilepsy) to evaluate the relevance of networks identified from scalp EEG data. Overall results obtained on simulated as well on real data indicated that the combination of the wMNE and the PLV methods leads to the most relevant networks as compared with the ground-truth (simulations) or with the intracerebrally-identified networks (patient data). Results are more specifically discussed hereafter.

*Methodological considerations*

The connectivity matrices were thresholded by keeping the nodes with highest strength values (strongest 1%). This procedure was used to standardize the comparison between all the combinations. We were aware about the effect of this threshold and we realized the comparative study using different threshold values. All threshold values were found to lead to the same differences between the method (inverse/connectivity) combinations.

In this paper, we have averaged the reconstructed sources within the same region of interest defined by the parcellation process based on Destrieux atlas. This approach was frequently used in the context of M/EEG source connectivity (de Pasquale et al. 2010; Fraschini et al. 2016; Hassan et al. 2015a). However, such an averaging procedure may increase the noise power since its computation is performed over sources that, with some probability, may not exhibit correlation (Brookes et al. 2014) where the need of alternative approaches such as flipping the sign of the sources in each ROIs before averaging the regional time series (Fraschini et al. 2016) or developing methods based on probabilistic maps, a widely approach used in the fMRI-based analysis, for instance.

Although EEG source connectivity reduced the problem of volume conduction as compared with scalp EEG connectivity, it does not yet provide a perfect solution. The volume conduction effect is a challenging issue when performing EEG/MEG inverse solution (Schoffelen and Gross 2009). In the connectivity context, the main effect of the volume conduction is the appearance of 'artificial' connections among close sources, a problem often referred to as 'source leakage'. The use of a high spatial resolution (high number of ROIs) may increase this problem. A few approaches have been proposed recently to deal with the source leakage by either normalizing the edges weights by the distance between the nodes or removing the edges between very close sources. Although, these approaches have some advantages, it was shown that, in most cases, they also remove 'real' connections (Schoffelen and Gross 2009). In this context, some connectivity methods such as PLV have been shown to reduce the volume conduction (Hipp et al. 2011). This may explain the good performance of this method. Indeed, in the double network scenario, PLV was able to detect the long-range connections between parietal and temporal networks.

Four inverse/connectivity algorithms were evaluated in this paper. It is worth mentioning that some other inverse algorithms like MUSIC-based and beamforming as well some connectivity measures such as power envelope correlation (O'Neill et al. 2015) were not included in this study. Moreover, we were focusing in this paper on evaluating different families of 'functional' connectivity methods regardless the directionality of these connections. Nevertheless, we consider that the analyses of the 'effective' connectivity methods that investigate the causality between brain regions may be of interest in the context of epilepsy (Coito et al. 2016; Coito et al. 2015). In this perspective, methods such as the granger causality, the transfer entropy could be added to expand this comparative study. In addition, all methods evaluated in the paper were bivariate, multivariate methods such as those based on the MVAR model were not included in our study. Different methodological questions appear when using MVAR-based approaches. First, the successful estimation of MVAR such as Partial Directed Coherence (PDC) or Directed Transfer Function (DTF) depends largely on the fitted MAR model, since all information is resulting from the estimated model parameters. In practice, this issue is difficult and directly related to the choice of an optimal model order and an optimal epoch length. Concerning the optimal model, most of the criteria were originally proposed for univariate AR modeling and no consensus was reported about multivariate ones. The second crucial question is how to choose the proper window size specially that MVAR model assumes that the underlying process is stationary, while neurophysiological activity are transient and may rapidly change their states representing high nonstationary behaviors (Pereda et al. 2005). Nevertheless the MVAR (when carefully applied) could provide complementary information not only about the link exists between two signals but also if one structure drive another of if there is feedback between these structures (Kuś et al. 2004). The directionality could be also defined as 'time-delay' between two regional time series which can be computed using linear or nonlinear correlation coefficients. As our main objective in this study was to compare inverse algorithms and 'functional' connectivity methods using same criteria (here we used similarity between reference and estimated undirected graphs), we didn't investigate the time-delays in the presented quantitative analysis. In addition this feature cannot be computed for all the selected methods (the case of the phase synchronization method for instance). We believe that the directionality, estimated from Granger causality or/and time delays, is indeed an interesting supplementary feature in the context of epileptic seizure propagation and will be a potential element for further analysis.

The head model used in this study was computed using the Boundary Element Method (BEM) with three layers (skin, skull and brain). This model was widely used in the context of M/EEG source estimation (Fuchs et al. 2007) as a compromise between computational cost and accuracy. Nevertheless, other methods exist to compute the head model such as the Finite Element Method (FEM) or adding other layers such as cerebrospinal fluid (CSF). These methods can possibly have effect of the resultant network (Cho et al. 2015). The evaluation of the above mentioned parameters/factors may be the topic of further investigation.

*Identification of interictal epileptic networks from scalp dense-EEG data*

A salient feature of epilepsy in general and epileptic networks is the increased synchronization among interconnected neuronal populations distributed over distant areas. This "hyper"- synchronization often leads to an increase of brain connectivity, not only during the transition to seizures but also during interictal periods, as shown in many studies based on intracranial recordings (see (Wendling et al. 2010) for review). In this context, the combination of the M/EEG source imaging with the functional connectivity measures has recently disclosed promising findings to identify pathological brain networks, at the cortical level (Dai et al. 2012; Lu et al. 2012; Malinowska et al. 2014; Song et al. 2013).

However, two factors seem to be crucial for reliable estimation of EEG source connectivity: i) the number of scalp electrodes and ii) the combination between the inverse solution algorithm and the functional connectivity measure. Concerning the number of electrodes, it was reported that the increase of the spatial resolution by using dense EEG may dramatically improve the accuracy of the source localization analysis (Michel and Murray 2012; Song et al. 2015). In addition, the use of dense EEG, as compared to classical montages (32 or 64 electrodes), is needed to accurately identify functional brain networks from scalp EEG (Hassan et al. 2014). To overcome this issue, dense-EEG (256 electrodes) recordings were used in this study. The main feature of this system is the excellent coverage of the subject's head by surface electrodes allowing for improved reconstruction of the cortical activity from non-invasive scalp measurements, as compared with more standard 32-128 electrode systems (Song et al. 2015). Regarding the combination of inverse/connectivity methods, most of reported studies have empirically selected a combination while this selection was shown to have a dramatic impacts on results, in term of identified network topology (Hassan et al. 2014). The present analysis brings further confirmation of this high variability observed when different inverse solutions and/or connectivity measures are being used in the pipeline leading to cortical networks from EEG signals.

A major advantage of the EEG source connectivity approach as presented here is that reconstructed sources and associated networks were identified for the whole brain. Then graph-based metrics (strength values) were computed to characterize the networks and the similarity index was used to compare the networks obtained from various method combinations. In addition, functional connectivity was applied directly to the reconstructed signals and not on derived components. In this regard, this study differs from (Malinowska et al. 2014) where connectivity was estimated on signals components obtained by ICA decomposition. Although the methodological issue of measuring connectivity between *independent* components still holds, a future interesting study will compare the results obtained from the ICA-based approach to those reported here from source connectivity.

*EEG source localization vs. functional connectivity*

Source localization methods have been widely applied to interictal epileptic spikes (Becker et al. 2014). The goal of these approaches is the localization of brain generators of epileptic activity from scalp recordings. A fundamental question that is addressed in this paper relates to the difference between the source connectivity and the source localization approach. This study indicated that the information extracted from dense-EEG recordings in both cases can differ dramatically. First, the connectivity is an additional step to the simple source reconstruction/localization. Second, the fundamental difference between both methods is that the source localization ignores all possible communications between brain regions. When performing source localization analysis, the sources with highest amplitude (averaged at given time period or computed at the instant of peak amplitude of the signal) are classically kept. However, to some extent (depending on threshold), this approach may neglect the possible contribution of "low energy" sources participating into the network activity.

Conversely, the hypothesis behind the network-based approach (typically when phase synchronization methods are used as connectivity measure) is that sources can be synchronized regardless their amplitude. To this extent, we believe that the network-based approach allows for revealing networks that are more specific to epileptic networks, as hyper-synchronization phenomena constitute a typical hallmark of such networks. An illustrative example in this paper is the poor involvements of the temporal lobe region when the sole source localization approach (in the case of cMEM) was applied while both parietal and temporal networks (as a priori introduced in the EEG generation model) are retrieved by the connectivity-based approach (cMEM/$h^2$). Note that we have averaged the source localization results in a time window of 50ms to cover the time delay of 30ms set in the model between the two brain regions. Different time window were used to avoid the effect of the window length. All tested windows (30ms, 40ms, 50ms and 60ms) showed similar observations i.e. the absence of the temporal sources (not shown here). The fact that epilepsy is considered as a network disease can explain the low performance of some of the inverse methods as these methods were originally developed to localize 'local' epileptic foci characterized by high-energy sources regardless the interrelationships between brain regions. Our results show that EEG source connectivity methods are more suited in the case of multi-focal epileptic zone. More generally, they support the recent tendency in brain disorder research which is the necessity to move from localizing 'pathological areas' to identifying 'altered networks' (Diessen et al. 2013; Fornito et al. 2015).

*Epilepsy as a network disorder*

There is increasing evidence that epileptic activity involves brain networks rather than a single well circumscribed region and that these dysfunctional networks contribute to both ictal and interictal activity (Bourien et al. 2005; Bourien et al. 2004; Coito et al. 2015; Engel Jr et al. 2013; Hipp et al. 2011). Functional connectivity was widely applied to depth-EEG data to predict seizures (Mormann et al. 2000) and identify epileptic networks in partial epilepsies (Bartolomei et al. 2001). These studies showed alterations of synchronization in brain networks during interictal to ictal transition (Wendling et al. 2003) as well as during seizures (Diessen et al. 2013; Jiruska et al. 2013; Schindler et al. 2008). Most of these studies were performed using invasively-recorded data in patient's candidate to surgery. Interestingly, our results show that pathological networks involved during epileptiform activity can also be identified from scalp EEG.

Indeed, we have evaluated the performance of a relatively new approach aiming at identifying epileptic brain networks from scalp EEG. The application of the method on real data showed the good performance of this method in term of network identified from scalp EEG as compared with those identified from intracerebral EEG. Note that the comparison was done only by computing $h^2$ between the intracerebral signals based on a large number of studies showing that $h^2$ is one of the most adapted metrics to compute functional connectivity between intracerebral recordings (Bettus et al. 2008; Wendling et al. 2010). Although it is obviously difficult to conclude on a single patient analysis, results showed good matching between scalp-EEG based networks and both the depth-EEG based networks and the expert judgment. Therefore, future work will consist in the application of the EEG source connectivity on a big database of real dense EEG data recorded from epileptic patients. In these patients candidate to surgery, we plan to use also intracerebral EEG signals as a reference to validate the identified networks. In addition, due to the excellent temporal resolution of the EEG, the dynamic behaviors of the epileptic networks will be also explored (Hassan et al. 2015a; Mheich et al. 2015b).

Finally, the capacity to describe epileptic activity not only according to the sites where epileptiform activity is generated but also according to the abnormal functional relationships between these sites can definitively improve the surgical approach. We speculate that in order to better understand and ultimately prevent seizures, it is essential to identify and then remove/disconnect pathological nodes of the network (exhibiting abnormal hyper-synchronization capability). The proposed method contributes to this aim and reported results constitute a first step toward the development of more efficient non-invasive diagnostic methods for clinical epileptology.

# Acknowledgements

This work has received a French government support granted to the CominLabs excellence laboratory and managed by the National Research Agency in the "Investing for the Future" program under reference ANR-10-LABX-07-01. It was also financed by the Rennes University Hospital (COREC Project named conneXion, 2012-14).

Bartolomei F, Wendling F, Bellanger J-J, Régis J, Chauvel P (2001) Neural networks involving the medial temporal structures in temporal lobe epilepsy Clinical neurophysiology 112:1746-1760

Baumgartner C et al. (1995) Propagation of interictal epileptic activity in temporal lobe epilepsy Neurology 45:118-122

Becker H et al. (2014) EEG extended source localization: tensor-based vs. conventional methods NeuroImage 96:143-157

Betti V, Della Penna S, de Pasquale F, Mantini D, Marzetti L, Romani GL, Corbetta M (2013) Natural scenes viewing alters the dynamics of functional connectivity in the human brain Neuron 79:782-797

Bettus G, Wendling F, Guye M, Valton L, Regis J, Chauvel P, Bartolomei F (2008) Enhanced EEG functional connectivity in mesial temporal lobe epilepsy Epilepsy Res 81:58-68 doi:10.1016/j.eplepsyres.2008.04.020

Bola M, Sabel BA (2015) Dynamic reorganization of brain functional networks during cognition NeuroImage 114:398-413

Bourien J, Bartolomei F, Bellanger JJ, Gavaret M, Chauvel P, Wendling F (2005) A method to identify reproducible subsets of co-activated structures during interictal spikes. Application to intracerebral EEG in temporal lobe epilepsy Clinical neurophysiology : official journal of the International Federation of Clinical Neurophysiology 116:443-455 doi:10.1016/j.clinph.2004.08.010

Bourien J, Bellanger JJ, Bartolomei F, Chauvel P, Wendling F (2004) Mining reproducible activation patterns in epileptic intracerebral EEG signals: application to interictal activity IEEE transactions on bio-medical engineering 51:304-315 doi:10.1109/TBME.2003.820397

Brookes MJ et al. (2014) Measuring temporal, spectral and spatial changes in electrophysiological brain network connectivity Neuroimage 91:282-299

Caparos M, Louis-Dorr V, Wendling F, Maillard L, Wolf D (2006) Automatic lateralization of temporal lobe epilepsy based on scalp EEG Clinical neurophysiology 117:2414-2423

Cho J-H, Vorwerk J, Wolters CH, Knösche TR (2015) Influence of the head model on EEG and MEG source connectivity analysis Neuroimage 110:60-77

Chowdhury RA, Lina JM, Kobayashi E, Grova C (2013) MEG source localization of spatially extended generators of epileptic activity: comparing entropic and hierarchical Bayesian approaches PloS one 8:e55969

Clarke C, Janday B (1989) The solution of the biomagnetic inverse problem by maximum statistical entropy Inverse Problems 5:483

Coito A et al. (2016) Altered directed functional connectivity in temporal lobe epilepsy in the absence of interictal spikes: A high density EEG study Epilepsia

Coito A et al. (2015) Dynamic directed interictal connectivity in left and right temporal lobe epilepsy Epilepsia 56:207-217

Cosandier-Rimélé D, Merlet I, Badier J-M, Chauvel P, Wendling F (2008) The neuronal sources of EEG: modeling of simultaneous scalp and intracerebral recordings in epilepsy NeuroImage 42:135-146

Dai Y, Zhang W, Dickens DL, He B (2012) Source connectivity analysis from MEG and its application to epilepsy source localization Brain topography 25:157-166

Dale AM, Liu AK, Fischl BR, Buckner RL, Belliveau JW, Lewine JD, Halgren E (2000) Dynamic statistical parametric mapping: combining fMRI and MEG for high-resolution imaging of cortical activity Neuron 26:55-67

David O, Cosmelli D, Hasboun D, Garnero L (2003) A multitrial analysis for revealing significant corticocortical networks in magnetoencephalography and electroencephalography Neuroimage 20:186-201

David O, Garnero L, Cosmelli D, Varela FJ (2002) Estimation of neural dynamics from MEG/EEG cortical current density maps: application to the reconstruction of large-scale cortical synchrony IEEE Transactions on Biomedical Engineering, 49:975-987

de Pasquale F et al. (2010) Temporal dynamics of spontaneous MEG activity in brain networks Proceedings of the National Academy of Sciences 107:6040-6045

Destrieux C, Fischl B, Dale A, Halgren E (2010) Automatic parcellation of human cortical gyri and sulci using standard anatomical nomenclature Neuroimage 53:1-15

Diessen E, Diederen SJ, Braun KP, Jansen FE, Stam CJ (2013) Functional and structural brain networks in epilepsy: what have we learned? Epilepsia 54:1855-1865

Ding L, Worrell GA, Lagerlund TD, He B (2007) Ictal source analysis: localization and imaging of causal interactions in humans Neuroimage 34:575-586

Ebersole J (1994) Non-invasive localization of the epileptogenic focus by EEG dipole modeling Acta Neurologica Scandinavica 89:20-28

Emerson RG, Turner CA, Pedley TA, Walczak TS, Forgione M (1995) Propagation patterns of temporal spikes Electroencephalography and clinical neurophysiology 94:338-348

Engel Jr J, Thompson PM, Stern JM, Staba RJ, Bragin A, Mody I (2013) Connectomics and epilepsy Current opinion in neurology 26:186

Fischl B (2012) FreeSurfer Neuroimage 62:774-781

Fornito A, Zalesky A, Breakspear M (2015) The connectomics of brain disorders Nature Reviews Neuroscience 16:159-172

Fraschini M, Demuru M, Crobe A, Marrosu F, Stam CJ, Hillebrand A (2016) The effect of epoch length on estimated EEG functional connectivity and brain network organisation Journal of Neural Engineering 13:036015

Fuchs M, Wagner M, Kastner J (2007) Development of volume conductor and source models to localize epileptic foci Journal of Clinical Neurophysiology 24:101-119

Gotman J (1983) Measurement of small time differences between EEG channels: method and application to epileptic seizure propagation Electroencephalography and clinical neurophysiology 56:501-514

Gotman J (1987) Interhemispheric interactions in seizures of focal onset: data from human intracranial recordings Electroencephalography and clinical neurophysiology 67:120-133

Gramfort A, Papadopoulo T, Olivi E, Clerc M (2010) OpenMEEG: opensource software for quasistatic bioelectromagnetics Biomed Eng Online 9:45

Hämäläinen M (2005) MNE software user's guide NMR Center, Mass General Hospital, Harvard University 58:59-75

Hämäläinen MS, Ilmoniemi R (1994) Interpreting magnetic fields of the brain: minimum norm estimates Medical & biological engineering & computing 32:35-42

Hassan M, Benquet P, Biraben A, Berrou C, Dufor O, Wendling F (2015a) Dynamic reorganization of functional brain networks during picture naming Cortex 73:276-288

Hassan M, Dufor O, Merlet I, Berrou C, Wendling F (2014) EEG Source Connectivity Analysis: From Dense Array Recordings to Brain Networks PloS one 9:e105041

Hassan M, Shamas M, Khalil M, El Falou W, Wendling F (2015b) EEGNET: An Open Source Tool for Analyzing and Visualizing M/EEG Connectome PloS one 10:e0138297

Hassan M, Wendling F (2015) Tracking dynamics of functional brain networks using dense EEG IRBM 36:324-328

Hipp JF, Engel AK, Siegel M (2011) Oscillatory synchronization in large-scale cortical networks predicts perception Neuron 69:387-396

Hoechstetter K, Bornfleth H, Weckesser D, Ille N, Berg P, Scherg M (2004) BESA source coherence: a new method to study cortical oscillatory coupling Brain topography 16:233-238

Jiruska P, de Curtis M, Jefferys JG, Schevon CA, Schiff SJ, Schindler K (2013) Synchronization and desynchronization in epilepsy: controversies and hypotheses The Journal of physiology 591:787-797 doi:10.1113/jphysiol.2012.239590

Kramer MA, Cash SS (2012) Epilepsy as a disorder of cortical network organization The Neuroscientist 18:360-372

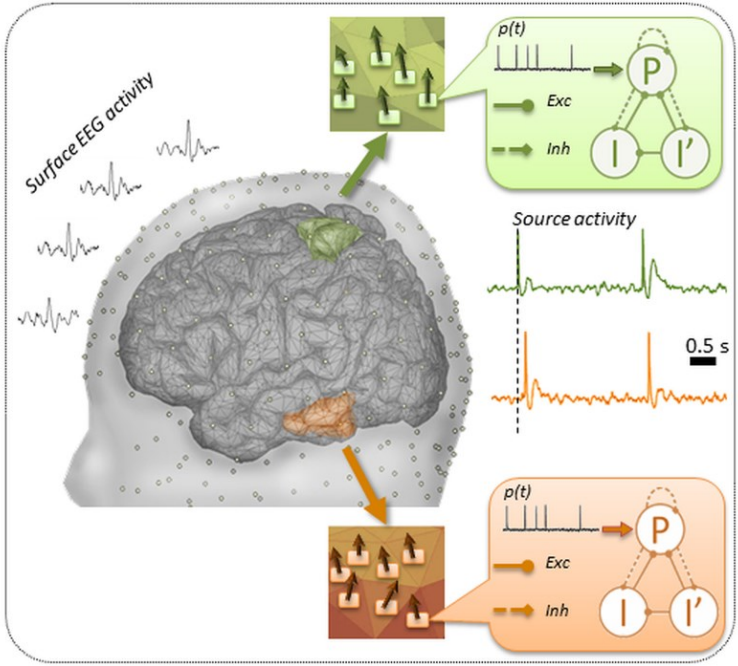
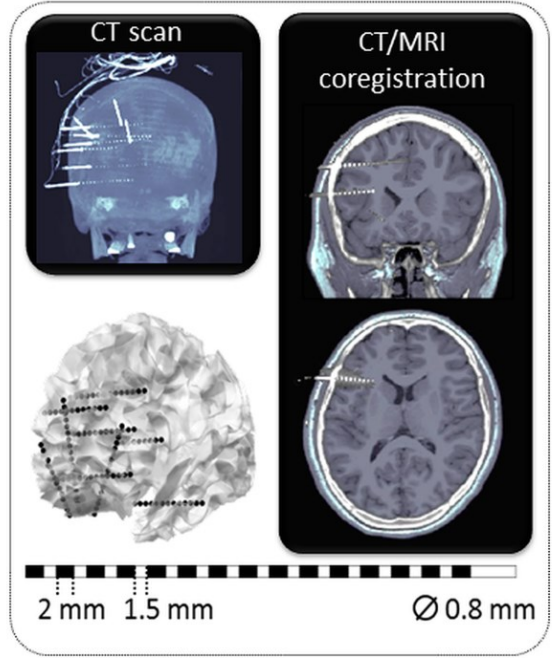
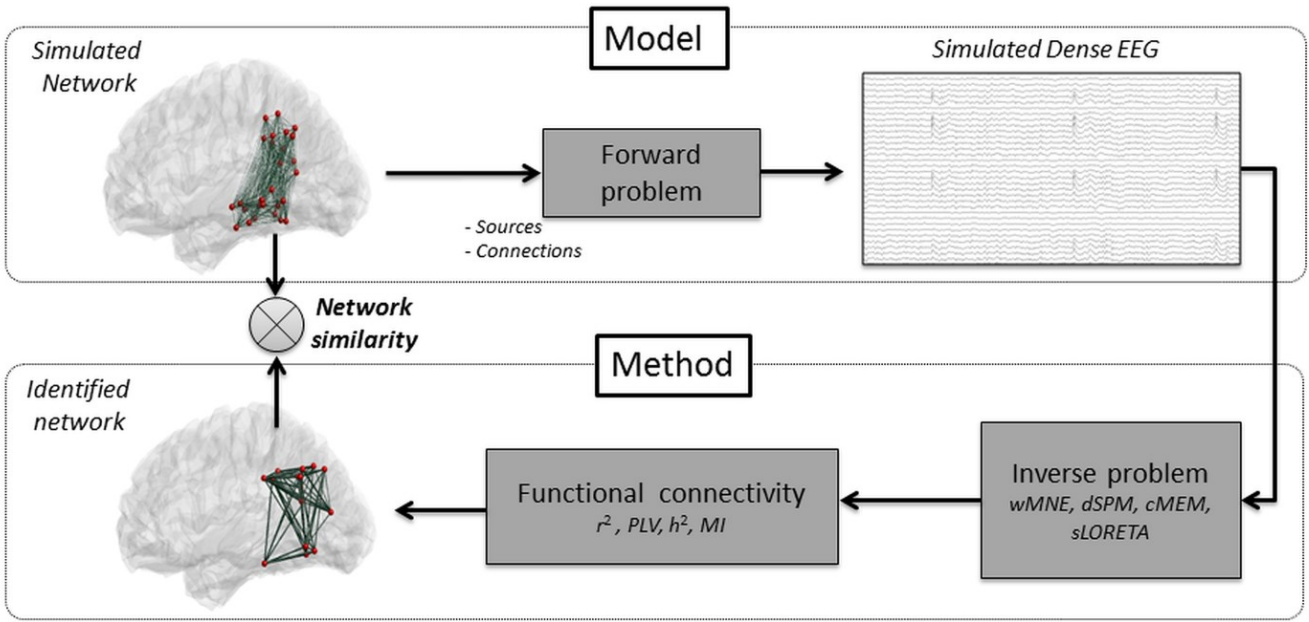

Figure 1: **Structure of the investigation**. A) **Simulated epileptic spikes**: Model used to generate epileptic spikes (see *Materials and methods/Data/Simulations section for detailed description*), B) **Identification of interictal epileptic network:** First, a network is generated by the model and considered as the 'ground truth'. By solving the forward problem, synthetic dense EEG data are generated. These signals are then used to solve the inverse problem in order to reconstruct the dynamics of sources using three different inverse solutions (wMNE, sLORETA, dSPM and cMEM). The statistical couplings are then computed between the reconstructed sources using three different methods ($r^2$, *PLV*, $h^2$ and MI). The identified network by each combination (inverse/connectivity) was then compared with the original network using a 'network similarity' algorithm [13] and C) **Intracerebral recordings:** The positions of the intracerebral SEEG signals used in the real application. The corrdinates of the electrode's contacts was obtianed by the CT/MRI coregistration. Abbreviations: wMNE: weighted Minimum Norm Estimate; sLORETA: Standardized low resolution brain electromagnetic tomography; dSPM: dynamical Statistical Parametric Mapping; cMEM: standard Maximum Entropy on the Mean; $r^2$: linear correlation coefficient; PLV: Phase Locking Value; $h^2$: nonlinear correlation coefficient; MI: mutual information, P: pyramidal cells, I: Inhibitory interneurons.

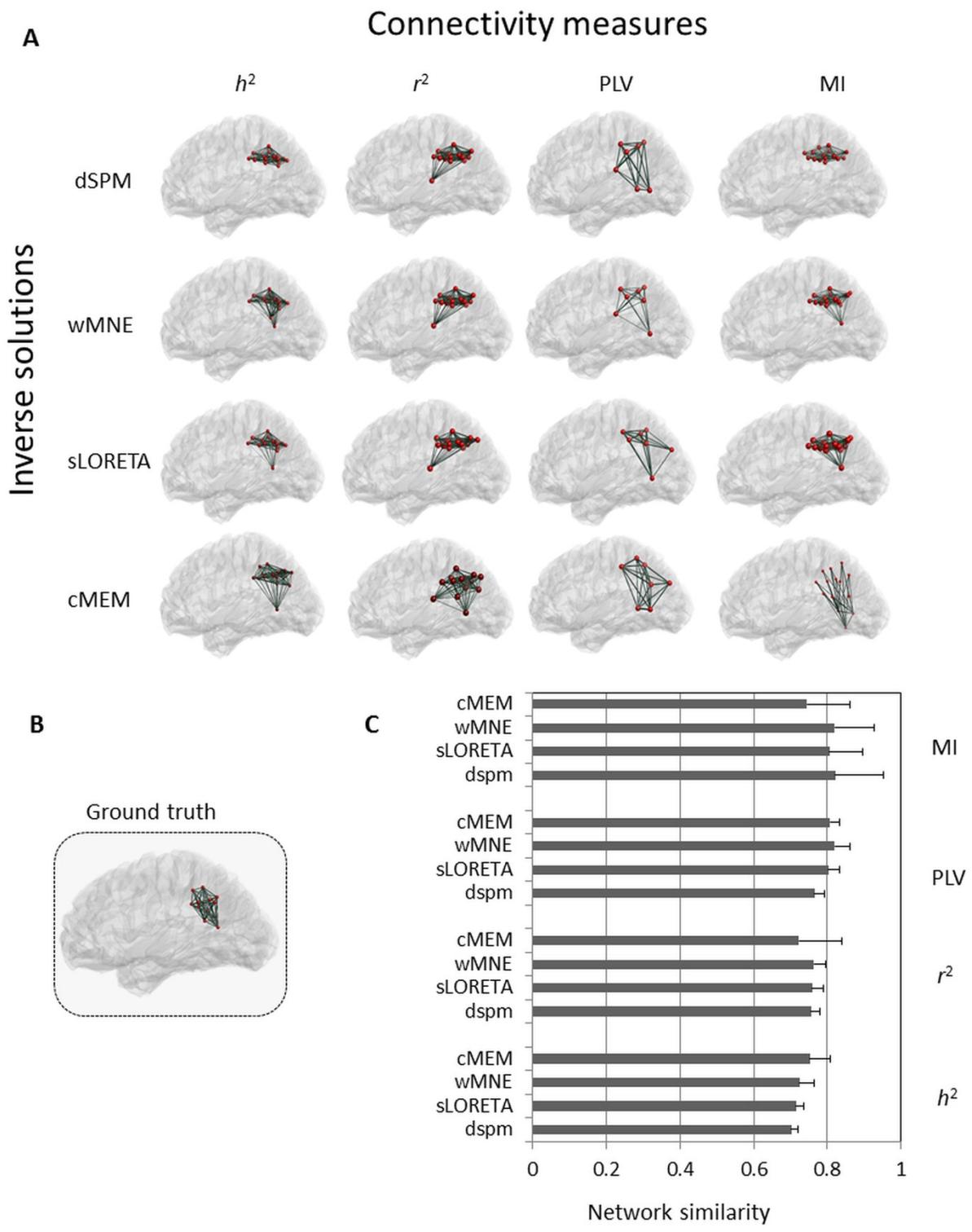

Figure 2: **One network scenario.** A) Brain networks obtained by using the different inverse and connectivity methods, B) The original network (ground truth) is shown and C) Values (mean ± standard deviation) of the similarity indices computed between the network identified by each combination and the model network.

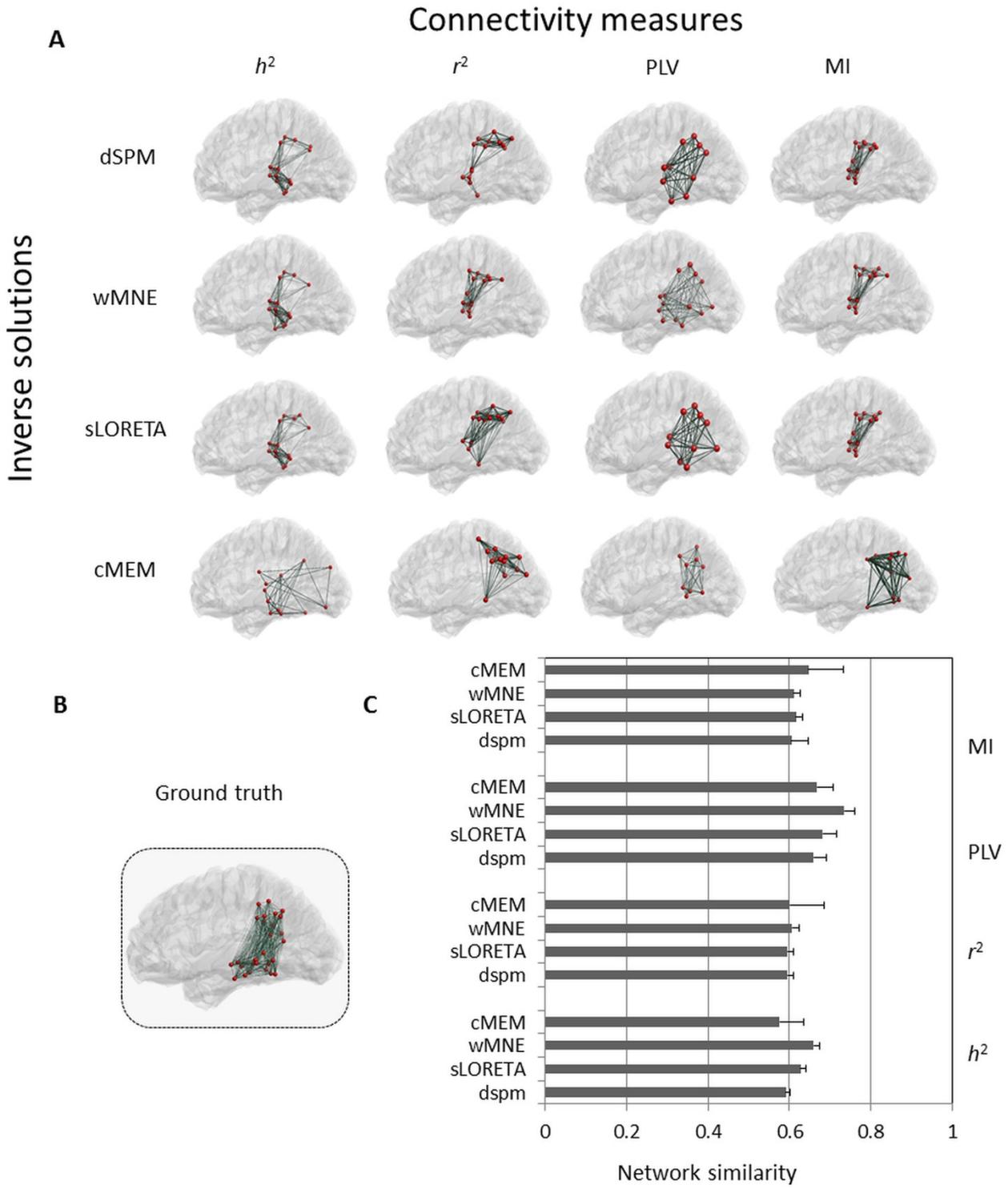

Figure 3: **Two networks scenario.** A) Brain networks obtained by using the different inverse and connectivity methods. B) The original network (ground truth) is shown and C) Values (mean ± standard deviation) of the similarity indices computed between the network identified by each combination and the model network.

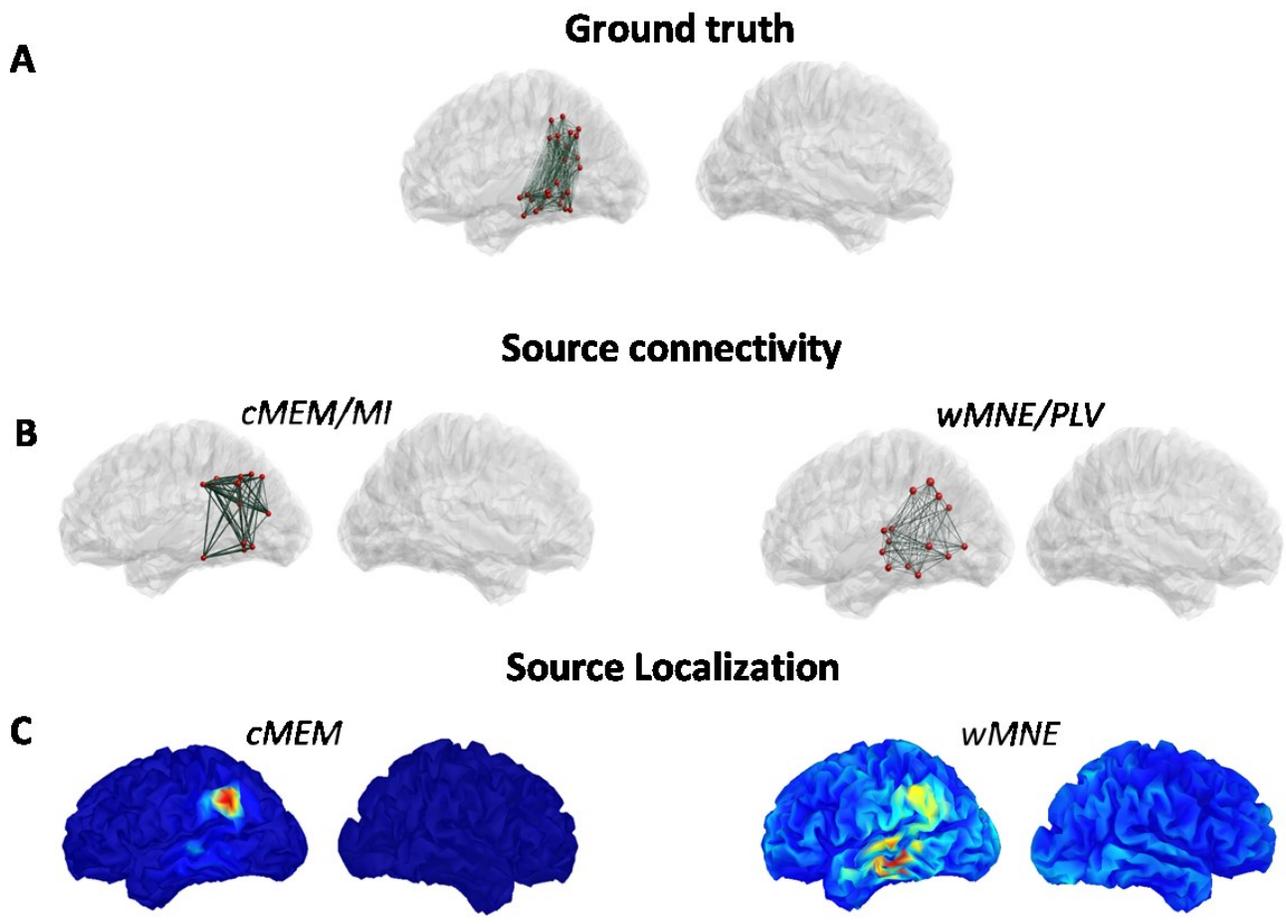

Figure 4: **Source localization vs. source connectivity**. A. the reference network. B. Results obtained by the network-based approach (cMEM/MI and wMNE/PLV) and C. Results obtained by the localization based approach (cMEM and wMNE) using same window of analysis. Results were averaged over a 50 ms interval around each of the spike peaks. Red color represents the sources with the highest energy.

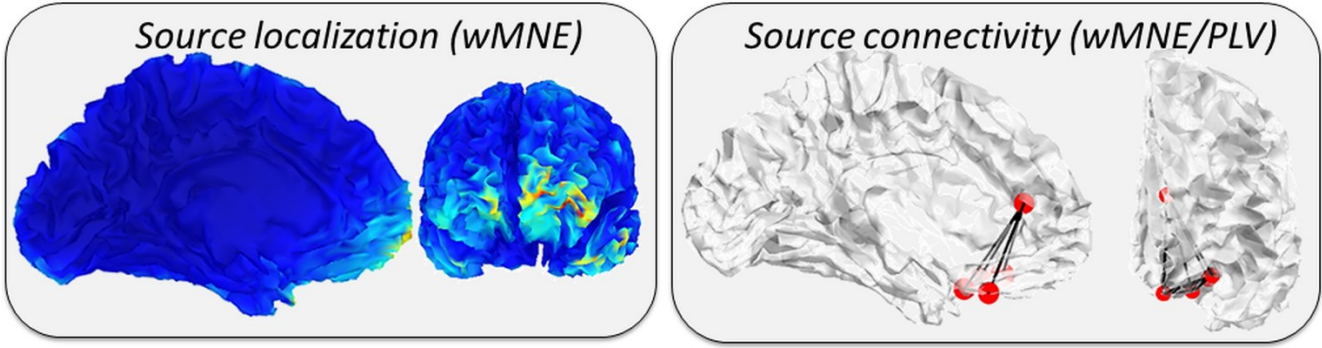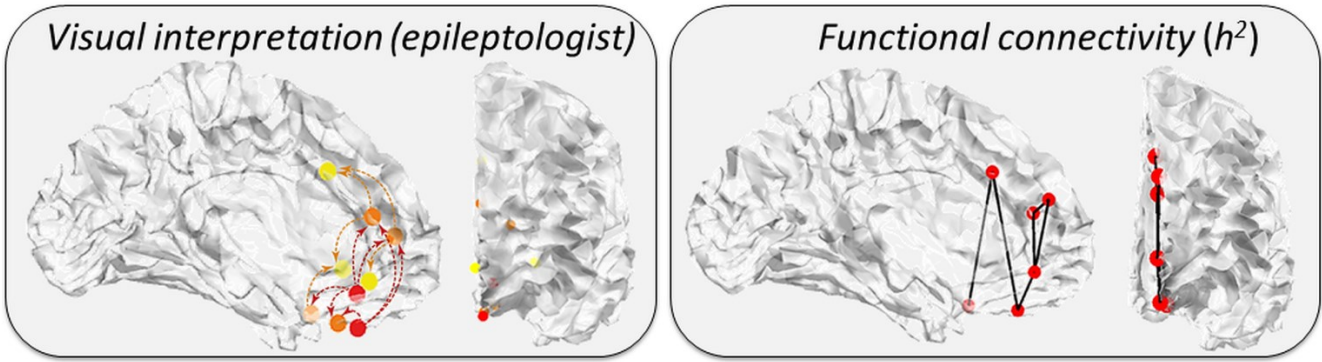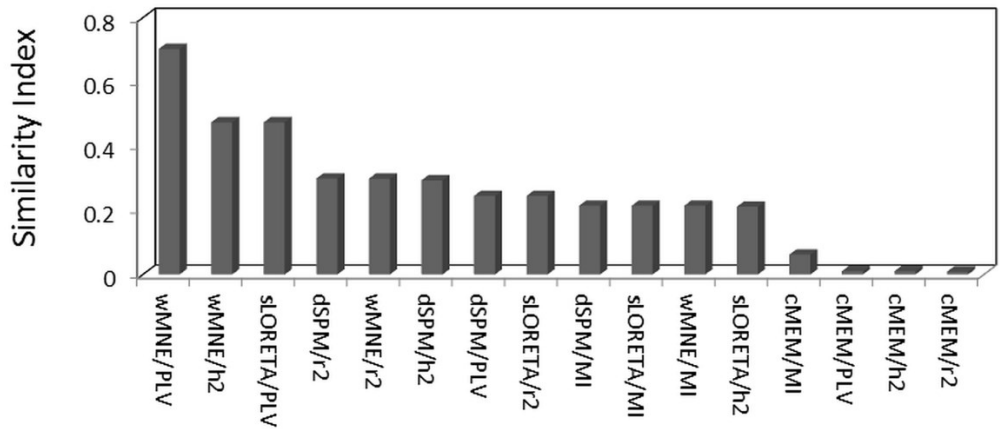

Figure 5: **Application on real data**. A) **Scalp EEG**: Results of the source localization approach using wMNE (right) and source connectivity using wMNE/PLV (left), B. **Intracerebral EEG**: Regions visually identified by the epileptologist (right) and the network obtained by computing the functional connectivity between the intracerebral EEG signals (left), C. **Similarity indices**: The SI values obtained between the network identified by each of the combination and the intracerebral-EEG-based network.